\documentclass[tradiabstract]{aa} 
\usepackage{graphicx}
\usepackage{txfonts}
\usepackage{natbib}
\begin{document}
%
   \title{Photospheric activity, rotation, and star-planet interaction of the planet-hosting star CoRoT-6\thanks{Based on observations obtained with CoRoT, a space project operated by the French Space Agency, CNES, with partecipation of the Science Programme of ESA, ESTEC/RSSD, Austria, Belgium, Brazil, Germany, and Spain.}} 
  
\authorrunning{A. F. Lanza et al.}
\titlerunning{Activity, rotation, and star-planet interaction of CoRoT-6}

   \author{A.~F.~Lanza\inst{1} \and A.~S.~Bonomo\inst{1,2} 
          \and I.~Pagano\inst{1} \and G.~Leto\inst{1} \and S.~Messina\inst{1} \and G.~Cutispoto\inst{1}
          \and C.~Moutou\inst{2}
          \and S.~Aigrain\inst{3} \and R.~Alonso\inst{4} \and P.~Barge\inst{2}, M.~Deleuil\inst{2}
         \and M.~Fridlund\inst{5}  
         \and A.~Silva-Valio\inst{6} 
         \and M.~Auvergne\inst{7} \and A.~Baglin\inst{7} 
       \and A.~Collier~Cameron\inst{8}  
          }

   \institute{INAF-Osservatorio Astrofisico di Catania, Via S.~Sofia, 78, 95123 Catania, Italy\\
              \email{nuccio.lanza@oact.inaf.it}
\and 
Laboratoire d'Astrophysique de Marseille (UMR 6110),
Technopole de Ch\^{a}teau-Gombert,
38 rue Fr\'ed\'eric Joliot-Curie,
13388 Marseille cedex 13, France     
\and Department of Physics, University of Oxford, Denys Wilkinson Building, Keble Road, Oxford OX1 3RH, United Kingdom
\and
Observatoire de Gen\`eve, Universit\'e de Gen\`eve, 51 Ch. des Maillettes, 1290, Sauverny, Switzerland  
\and
Research and Scientific Support Department, European Space Agency, Keplerlaan 1, 2200AG, Noordwijk, The Netherlands 
\and
CRAAM, Mackenzie University, Rua de Consola\c{c}\~{a}o, 896, 01302-907, S\~{a}o Paulo, Brazil
\and 
    LESIA, CNRS UMR 8109, Observatoire de Paris, 5 place J. Janssen, 92195 Meudon, France       
    \and 
    School of Physics and Astronomy, University of St. Andrews, 
    North Haugh, St Andrews, Fife  KY16 9SS, Scotland   
    }

   \date{Received 21 June 2010 ; accepted 16 September 2010}

 
  \abstract
   {The CoRoT satellite has recently discovered a hot Jupiter that transits across the disc of a F9 main-sequence star called CoRoT-6 with a period of 8.886 days. }
   {We model the photospheric activity of the star and use the maps of the active regions to study stellar  differential rotation and the star-planet interaction.   }
   {We apply a maximum entropy spot model  to fit the  optical modulation as observed by CoRoT during a uninterrupted interval of $\sim 140$ days. Photospheric active regions are assumed to consist of spots and faculae  in a fixed proportion with solar-like contrasts. }
   {Individual active regions have lifetimes up to $30-40$ days. Most of them form and decay within five active longitudes whose  different migration rates are attributed to the stellar differential rotation for which a lower limit of $\Delta \Omega / \Omega = 0.12 \pm 0.02$ is obtained. Several active regions show a maximum of activity at a longitude lagging the subplanetary point by $\sim 200^{\circ}$ with the probability of a chance occurrence  being smaller than 1 percent. } 
  {Our spot modelling  indicates that the photospheric activity of CoRoT-6 could be partially modulated by some kind of star-planet magnetic interaction, while an interaction related to tides is highly unlikely because of the  weakness of the tidal force. 
    }


   \maketitle
%

\section{Introduction}

CoRoT (Convection, Rotation and Transits) is a photometric space experiment devoted to asteroseismology and the search for extrasolar planets by the method of transits \citep{Baglinetal06}. It has recently discovered a hot Jupiter with an orbital period of 8.886 days and a mass of $2.96 \pm 0.34$ M$_{\rm J}$ orbiting a F9 main-sequence star dubbed \object{CoRoT-6} \citep{Fridlundetal10}. The star is magnetically active and shows a rotational modulation of its optical flux with a relative  amplitude of $\sim 2.7$ percent and a period of $\sim 6.4$ days.

We study the photospheric activity of CoRoT-6 by applying the spot modelling approach already used for the high-precision light curves of \object{CoRoT-2} 
\citep{Lanzaetal09a}, \object{CoRoT-4} \citep{Lanzaetal09b}, and \object{CoRoT-7} \citep{Lanzaetal10}. It allows us to map the longitudinal distribution of photospheric active regions and trace their evolution during the $\sim 140$ days of CoRoT observations. The different rotation rates of active regions  allow us to estimate a lower limit for the amplitude of  the differential rotation of CoRoT-6. We also use the maps to look for signs of star-planet interaction and find some statistical evidence of active regions lagging the subplanetary point by $\sim 170^{\circ}-200^{\circ}$ during some time intervals.  The case for a star-planet  interaction in other systems has been recently reviewed by \citet{Lanza08} and \citet{Shkolniketal09}.

\section{Observations}
\label{observations}

CoRoT-6 has been observed during the second long run of CoRoT toward the galactic centre from 15 April to 7~September  2008. Since the star is relatively bright ($ V =13.91$ mag), the time sampling is 32~s from the beginning of the observations. CoRoT performs aperture photometry with a fixed mask \citep[see Fig.~3 in ][]{Fridlundetal10}. The contaminating  flux from background stars falling inside the mask amounts to $2.8 \pm 0.7$ percent and produces a dilution of the flux of CoRoT-6  that has been corrected by subtracting the 2.8 percent of its median flux value. The flux inside the star's mask is split along detector column boundaries into broad-band red, green, and blue channels. To increase the signal-to-noise ratio and reduce systematic drifts present in individual channels, we sum up the flux in the red, green, and blue channels to obtain a  light curve in a spectral range extending from 300 to 1100 nm. 

The observations and data processing are described by \citet{Fridlundetal10}, to whom we refer the reader for details. The reduction pipeline applies corrections for the sky background and the pointing jitter of the satellite, which is particularly relevant during  ingress and  egress from the Earth shadow. Measurements during the crossing of the South Atlantic Anomaly of the Earth's magnetic field, which amounts to about $7-9$ percent of each satellite orbit, are discarded. High-frequency spikes due to cosmic ray hits and outliers are removed by applying a 5-point running mean. The final duty cycle of the 32-s observations is $\geq 90$ percent for the so-called N2 data time series that are accessible through the CoRoT Public Data Archive at IAS ({http://idoc-corot.ias.u-psud.fr/}). More information on the instrument, its operation, and performance can be found in \citet{Auvergneetal09}.

The transits are removed from the light curve using the ephemeris of \citet{Fridlundetal10} and the out-of-transit data are binned by computing average flux values along each orbital period of the satellite (6184~s). This has the advantage of removing tiny systematic variations associated with the orbital motion of the satellite, still preserving the information on magnetic-induced variability \citep[cf. ][]{Alonsoetal08,Auvergneetal09}.  In such a way, we obtain a  light curve consisting of 1993 mean points  ranging from HJD 2454572.5075 to 2454717.4519, i.e., with a duration of 144.9443 days. The average standard error of the mean points is $ 2.41 \times 10^{-4}$ in relative flux units. We remove the long-term decrease of the flux, which may be  due to the aging of the detector, by fitting a straight line to the data as in the case of CoRoT-4 \citep[cf. ][]{Aigrainetal08,Lanzaetal09b}. 
The  maximum  flux in the de-trended binned time series at HJD~2454577.5245 is adopted as a reference unit level corresponding to the unspotted star, since the true value of the unspotted flux is unknown.

\section{Light curve  modelling}

\label{spotmodel}

The reconstruction of the surface brightness distribution from the rotational modulation of the stellar flux is an ill-posed problem, because the variation of the flux vs. rotational phase contains only information  on the distribution of the brightness inhomogeneities vs. longitude. The integration over the stellar disc effectively cancels any latitudinal information, particularly when the inclination of the rotation axis along the line of sight is close to $90^{\circ}$, as it is assumed in the present case \citep[see Sect.~\ref{model_param} and ][]{Lanzaetal09a}. Therefore, we need to include a priori information in the light curve inversion process to obtain a unique and stable map. This is  done by computing a maximum entropy (hereafter ME) map, which has been proven to successfully reproduce active region distributions and area variations in the case of the Sun \citep[cf. ][]{Lanzaetal07}. Our approach has  been compared to other spot modelling methods by \citet{Mosseretal09} and
\citet{Huberetal10}. 
{ The occultations of the stellar disc by the planet during transits can be used to resolve the fine structure of the spot pattern and study its  evolution \citep[e.g.,][]{SilvaValioetal10}. We plan to apply this approach  to CoRoT-6 in a forthcoming paper (Silva-Valio et al., in prep.).}

In our model, the star is subdivided into 200 surface elements, namely  200  squares of side $18^{\circ}$, with  each element containing unperturbed photosphere, dark spots, and facular areas. The fraction of an element covered by dark spots is indicated by the filling factor $f$,  the fractional  area of the faculae is $Qf$, and the fractional area of the unperturbed photosphere is $1-(Q+1)f$. 
The contribution to the stellar flux coming from the $k$-th surface element at the time $t_{j}$, where $j=1,..., N$   is an index numbering the $N$ points along the light curve, is given by:
\begin{eqnarray}
\Delta F_{kj} & = & I_{0}(\mu_{kj}) \left\{ 1-(Q+1)f + c_{\rm s} f +  \right. \nonumber \\
  & & \left.  Q f [1+c_{\rm f} (1 -\mu_{kj})] \right\} A_{k} \mu_{kj} {w}(\mu_{kj}),
\label{delta_flux}
\end{eqnarray}
where $I_{0}$ is the specific intensity in the continuum of the unperturbed photosphere at the isophotal wavelength of the observations, $c_{\rm s}$ and $c_{\rm f}$ are the spot and facular contrasts, respectively \citep[cf. ][]{Lanzaetal04}, $A_{k}$ is the area of the $k$-th surface element,
\begin{equation}
 {w} (\mu_{kj}) = \left\{ \begin{array}{ll} 
                      1  & \mbox{if $\mu_{kj} \geq 0$}  \\
                      0 & \mbox{if $\mu_{kj} < 0$ }
                              \end{array} \right. 
\end{equation}
is its visibility, and 
\begin{equation}
\mu_{kj} \equiv \cos \psi_{kj} = \sin i \sin \theta_{k} \cos [\ell_{k} + \Omega (t_{j}-t_{0})] + \cos i \cos \theta_{k},
\label{mu}
\end{equation}
is the {cosine} of the angle $\psi_{kj}$ between the normal to the surface element and the direction of the observer, with $i$ being the inclination of the stellar rotation axis along the line of sight, $\theta_{k}$ the colatitude and $\ell_{k}$ the longitude of the $k$-th surface element; $\Omega$ {denotes} the angular velocity of rotation of the star ($\Omega \equiv 2 \pi / P_{\rm rot}$, with $P_{\rm rot}$ the stellar rotation period), and $t_{0}$ the initial time. The specific intensity in the continuum varies according to a quadratic limb-darkening law, as adopted by \citet{Lanzaetal03} for the case of the Sun, viz. $I_{0} \propto a_{\rm p} + b_{\rm p} \mu + c_{\rm p} \mu^{2}$. The stellar flux computed at the time $t_{j}$ is then: $F(t_{j}) = \sum_{k} \Delta F_{kj}$. To warrant a relative precision of the order of $10^{-5}$ in the computation of the flux $F$, each surface element is further subdivided into $1^{\circ} \times 1^{\circ}$-elements  and their contributions, calculated according to Eq.~(\ref{delta_flux}), are summed up at each given time to compute the contribution of the $18^{\circ} \times 18^{\circ}$ surface element to which they belong.  

We fit the light curve by varying the value of $f$ over the surface of the star, while $Q$ is held constant. Even fixing the rotation period, the inclination, and the spot and facular contrasts \citep[see ][ for details]{Lanzaetal07}, the model has 200 free parameters and suffers from  non-uniqueness and instability. To find a unique and stable spot map, we apply ME regularization by minimizing a functional $Z$, which is a linear combination of the $\chi^{2}$ and  the entropy functional $S$; i.e.,
\begin{equation}
Z = \chi^{2} ({\vec f}) - \lambda S ({\vec f}),
\end{equation}
where ${\vec f}$ is the vector of the filling factors of the surface elements, $\lambda > 0$   a Lagrangian multiplier determining the trade-off between light curve fitting and regularization; the expression for $S$ is given in \citet{Lanzaetal98}.  The entropy functional $S$  attains its maximum value when the star is {free of spots}. Therefore, by increasing the Lagrangian multiplier $\lambda$, we increase the weight of $S$ in the model and the area of the spots  is progressively reduced.
This gives rise to systematically negative residuals between the observations and the best-fit model when
$\lambda > 0$. The optimal value of $\lambda$ depends on the information content of the light curve, which in turn depends on the ratio of the amplitude of its rotational modulation to the average errors of its  points. In the case of CoRoT-6, the amplitude of the  rotational modulation is $\sim 0.027$, while the average errors of the  points is $\sim 2.4 \times 10^{-4}$ in relative flux units, giving a signal-to-noise ratio of $\sim 110 $. This is sufficient to adopt the same regularization criterion applied in the case of CoRoT-2 \citep[see ][]{Lanzaetal09a}, {iterating the value of $\lambda$ until the condition $|\mu_{\rm reg}| \simeq \epsilon_{0}$ is met, where $\mu_{\rm reg}$ is the mean of the residuals and $\epsilon_{0} \equiv \sigma_{0}/\sqrt{N}$ their standard error, i.e., the ratio of their standard deviation $\sigma_{0}$ in the case of the unregularized best fit (i.e., for $\lambda = 0$) to the square root of the number $N$ of  points in each fitted subset of the light curve having a duration  $\Delta t_{\rm f}$ (see below).} 

In the case of the Sun, by assuming a fixed distribution of the filling factor, it is possible to obtain a good fit of the irradiance changes only for a limited time interval $\Delta t_{\rm f}$, not exceeding 14 days, which is the lifetime of the largest sunspot groups dominating the irradiance variation. In the case of other active stars, the value of        $\Delta t_{\rm f}$ must be determined from the observations themselves, looking for the longest time interval that allows  a good fit with the applied model (see Sect.~\ref{model_param}). 

The optimal values of the spot and facular contrasts and of the facular-to-spotted area ratio $Q$ in stellar active regions are  unknown a priori. In our model the facular contrast $c_{\rm f}$ and the parameter $Q$ enter as the product $c_{\rm f} Q$, so we can fix $c_{\rm f}$ and vary $Q$, estimating its best value 
 by minimizing the $\chi^{2}$ of the model, as shown in Sect.~\ref{model_param}. Since the number of free parameters of the ME model is large, for this specific application we make use of the model of \citet{Lanzaetal03}, which fits the light curve by assuming only three active regions to model the rotational modulation of the flux plus a uniformly distributed background to account for the variations of  the mean light level. This procedure is the same used to fix the value of $Q$ in the cases of CoRoT-2, CoRoT-4, and CoRoT-7 
\citep[cf. ][]{Lanzaetal09a,Lanzaetal09b,Lanzaetal10}.  

We  assume an inclination of the rotation axis of CoRoT-6 of $ i \simeq 89^{\circ}$ in most of our models (see Sect.~\ref{model_param}). Since the information on spot latitude that can be extracted from the rotational modulation of the flux for such a high inclination is negligible, the ME regularization virtually puts all the spots around the sub-observer latitude (i.e., $90^{\circ} -i \approx 0^{\circ}$) to minimize their area and maximize the entropy. Therefore, we are limited to mapping  only the distribution of the active regions vs. longitude, which can be done with a resolution better than   $\sim 50^{\circ}$ \citep[cf. ][]{Lanzaetal07,Lanzaetal09a}. Our ignorance of the true facular contribution  may lead  to systematic errors in the active region longitudes derived by our model because faculae produce an increase of the flux when they are close to the limb, leading to a systematic shift of the longitudes of the active regions used to reproduce the observed flux modulation,   as discussed  by \citet{Lanzaetal07} {for} the case of the Sun and illustrated by \citet[][ cf.~ Figs.~4 and 5]{Lanzaetal09a} for CoRoT-2.

\section{Model parameters}
\label{model_param}

The fundamental stellar parameters are taken from \citet{Fridlundetal10} and are listed in 
Table~\ref{model_param_table}.  The limb-darkening parameters $a_{\rm p}$, $b_{\rm p}$, and $c_{\rm p}$ (cf. Sect.~\ref{spotmodel})  have been derived from \citet{Kurucz00} model atmospheres for $T_{\rm eff} = 6090$~K, $\log g = 4.44$~(cm~s$^{-2}$) and a metal abundance  $[M/H]= -0.2$, by adopting the CoRoT white-band transmission profile given by  \citet{Auvergneetal09}. The limb-darkening parameters $u_{+}=0.586 \pm 0.068$ and $u_{-}= -0.12 \pm 0.13$ obtained by fitting the transit cannot be used to constrain the above parameters because of their large errors (especially for $u_{-}$). However, they are compatible with those derived  from Kurucz's models that yield   $u_{+}=0.646$ and $u_{-}=-0.110$. 

The rotation period adopted for our spot modelling  has been derived from a Lomb-Scargle periodogram analysis of the light curve giving $P_{\rm rot} = 6.35 \pm 0.29$ days \citep{Scargle82}. The uncertainty of the period  comes from the total extension of the time series and represents an upper limit. As we shall see below, our models show that the starspots have a remarkable differential rotation, which may contribute to an uncertainty  of the stellar rotation period of $\approx 10$ percent, i.e., of $\pm \, 0.3$ days, comparable with the above upper limit (cf. Sect.~\ref{spot_model_res}). 

\begin{table}
\noindent 
\caption{Parameters adopted for the light curve  modelling of CoRoT-6.}
\begin{tabular}{lrr}
\hline
 & & \\
Parameter &  & Ref.$^{a}$\\
 & & \\ 
\hline
 & &  \\
Star Mass ($M_{\odot}$) & 1.05 & F10  \\
Star Radius ($R_{\odot}$) & 1.025 & F10  \\
$T_{\rm eff}$ (K) & 6090 &  F10 \\
$\log g$ (cm s$^{-2}$) & 4.44 & F10 \\ 
$a_{\rm p}$ & 0.340 & La10 \\
$b_{\rm p}$ & 1.024 & La10 \\
$c_{\rm p}$ & -0.378 & La10 \\ 
$P_{\rm rot}$ (days) & 6.35 & La10 \\
$\epsilon$ & $1.71 \times 10^{-4}$ & La10 \\ 
Inclination (deg) & 89.07 & F10  \\
$c_{\rm s}$  & 0.677 & La10 \\
$c_{\rm f}$  & 0.115 & L04 \\ 
$Q$ & 1.5  & La10 \\ 
$\Delta t_{\rm f}$ (days) & 6.15, 7.247 & La10 \\ 
& &   \\
\hline
\label{model_param_table}
\end{tabular}
~\\
$^{a}$ References: F10: \citet{Fridlundetal10};  L04: \citet{Lanzaetal04};  La10: present study.  
\end{table}

The polar flattening of the star due to the centrifugal potential  is computed in the Roche approximation with a rotation period of 6.35 days. The relative difference between the equatorial and the polar radii  $\epsilon$ is  $1.71 \times 10^{-4}$ which induces a completely negligible relative flux variation of $\approx 10^{-6}$  for a spot coverage of $\sim 2$ percent as a consequence of the gravity darkening of the equatorial region of the star. 

The inclination of the stellar rotation axis has not been constrained yet through the observation of the Rossiter-McLaughlin effect. Nevertheless, we assume that the stellar spin is normal to the orbital plane of the transiting planet, i.e., has an inclination of $89\, \fdg 07  \pm 0\, \fdg 30$ from the line of sight, unless otherwise specified. This is compatible with the measured $v \sin i =8.0 \pm 1.0$ km~s$^{-1}$, the above rotation period, and a radius estimate of $1.025 \pm 0.026$ R$_{\odot}$ from stellar evolutionary models, that yield $ \sin i = 0.98$ \citep[cf. ][]{Fridlundetal10}. Taking into account their  errors, the minimum inclination still compatible with those stellar parameters is $\sim 50^{\circ}$. 

The maximum time interval that our model can accurately fit with a fixed distribution of active regions $\Delta t_{\rm f}$ has been determined by dividing the total interval, $T= 144.9443$ days, into $N_{\rm f}$ equal segments, i.e., $\Delta t_{\rm f} = T/N_{\rm f}$, and looking for the minimum value of $N_{\rm f}$ that allows us a good fit of the light curve, as measured by the $\chi^{2}$ statistics. We find that for $N_{\rm f} < 20$ the quality of the best fit degrades significantly with respect to $N_{\rm f} \geq 20$, owing to a  substantial evolution of the pattern of surface brightness inhomogeneities. Therefore, we adopt   $\Delta t_{\rm f} = 7.2472$ days as the maximum time interval to be fitted with a fixed distribution of surface active regions in order to  estimate the best value of the parameter $Q$ (see below).  For the spot modelling in Sects.~\ref{light_curve_model} and ~\ref{spot_model_res}, we shall  consider the light curve interval from HJD~2454576.0 to 2454717.4519 because the light modulation is not well fitted during the first three days 
owing to the rapid evolution of some active regions (see Sect.~\ref{light_curve_model}), and adopt $N=23$ 
 corresponding to $\Delta t_{\rm f} = 6.15$ days, which provides a better time resolution to study the evolution of the spot pattern during the intervals with  faster modifications. 

To compute the spot contrast, we adopt the same mean temperature difference as derived for sunspot groups from their bolometric contrast, i.e., 560~K \citep{Chapmanetal94}. In other words, we assume a spot effective temperature of $ 5530$~K, yielding a contrast $c_{\rm s} = 0.677$ in the CoRoT passband \citep[cf. ][]{Lanzaetal07}.
A different spot contrast changes the absolute spot coverages, but does not significantly affect their longitudes and their time evolution, as discussed in detail by \citet{Lanzaetal09a}. The facular contrast is assumed to be solar-like with $c_{\rm f} = 0.115$ \citep{Lanzaetal04}. 

\begin{figure}[]
\centerline{
\includegraphics[width=8cm,height=6cm]{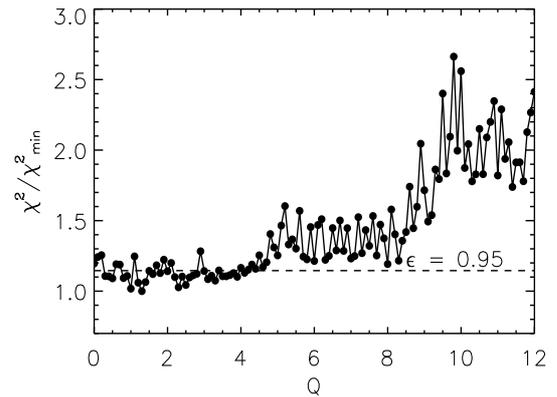}} 
\caption{
The ratio of the $\chi^{2}$ of the composite best fit of the entire time series (from HJD 2454572.5075 to 2454717.4519) to its minimum value vs. the parameter $Q$, i.e., the ratio of the {facular area  to  the cool spot area} in active regions. The horizontal dashed line indicates the 95 percent confidence level for $\chi^{2}/\chi_{\rm min}^{2}$, determining the interval of acceptable $Q$ values.
}
\label{qratio}
\end{figure}

The best value of the area ratio $Q$ of the faculae to the spots in each active region has been estimated by means of the model of \citet[][ cf. Sect.~\ref{spotmodel}]{Lanzaetal03}. In Fig.~\ref{qratio}, we plot the ratio $\chi^{2}/ \chi^{2}_{\rm min}$ of the 
total $\chi^{2}$ of the composite best fit of the entire time series to 
its minimum value $\chi^{2}_{\rm min}$, versus $Q$, and indicate the 95 percent confidence level as derived from the F-statistics \citep[e.g., ][]{Lamptonetal76}. Choosing $\Delta t_{\rm f} = 7.2472$ days, we fit the rotational modulation of the active regions for the longest time interval during which they remain stable, modelling both the flux increase due to the facular component when an active region is close to the limb as well as the flux decrease due to the dark spots when the same region transits across the central meridian of the disc. In such a way, a measure of the relative facular and spot contributions can be obtained leading to a fairly accurate estimate of $Q$. 

The best value of $Q$ turns out to be $Q \simeq 1.5$, with an acceptable range extending from~$0$~to~$\sim 4$. 
The $\chi^{2}$  fluctuates owing to the numerical errors occurring during its  minimization when a given best fit is computed. Therefore, it is its overall variation versus $Q$ that is relevant here.  
In the case of the Sun, the best value of $Q$ is~$9$ \citep{Lanzaetal07}. Thus our result indicates a lower relative contribution of the faculae to the  light variation  of CoRoT-6 than in the solar case. 
The relative amplitude of the rotational modulation of the star is $\sim 0.027$, showing that CoRoT-6 is more active than the Sun, for which a relative variation of $\sim 0.0035$ is observed at the maximum of the 11-yr cycle. 
This may explain the reduced facular contribution to the light variations of CoRoT-6, as suggested by \citet{Radicketal98} and \citet{Lockwoodetal07} who find a smaller facular signature in stars  more active than the Sun. 

The use of the chromatic information of the CoRoT light curves to estimate the spot and facular contrasts and filling factors is made impossible by our ignorance of the unperturbed stellar flux levels in the different colour channels which are needed to disentangle the flux perturbations due to spots and faculae, respectively. The continuous variations of the observed fluxes do not allow us to fix such reference levels so that we cannot unambiguously attribute a given flux modulation to cool spots or bright faculae. 
Moreover, the oscillations of the fluxes in the individual colour channels 
 on timescales comparable with the rotation period and the lack of precise  passband profiles, make the extraction of information on spots and faculae unfeasible.

\begin{figure}[]
\centerline{
\includegraphics[width=8cm,height=8cm,angle=90]{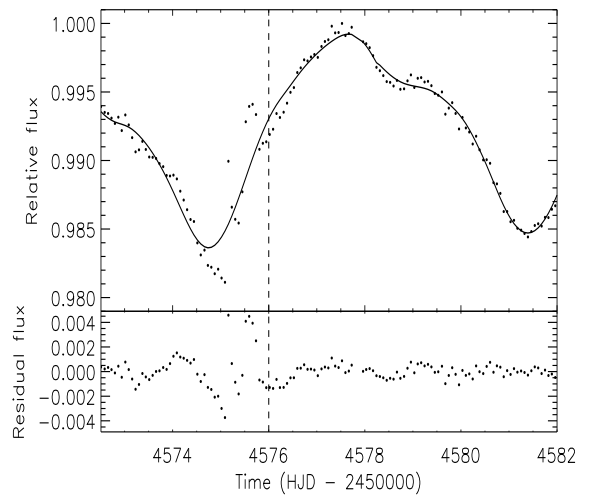}} 
\caption{{\it Upper panel:} The out-of-transit light curve of CoRoT-6 (dots) in the time interval between  HJD~2454572.5075 and 2454582.0 and its unregularized  best fit  for a facular-to-spotted area ratio of $Q=1.5$ (solid line). The flux is normalized to the maximum  at HJD~2454577.5245. {\it Lower panel:} The corresponding residuals. Note the greater residuals  before HJD~2454576.0 (marked by the vertical dashed line) owing to the rapid evolution of some active regions. 
}
\label{initial_segment}
\end{figure}
\begin{figure*}[]
\centerline{
\includegraphics[width=9cm,height=20cm,angle=90]{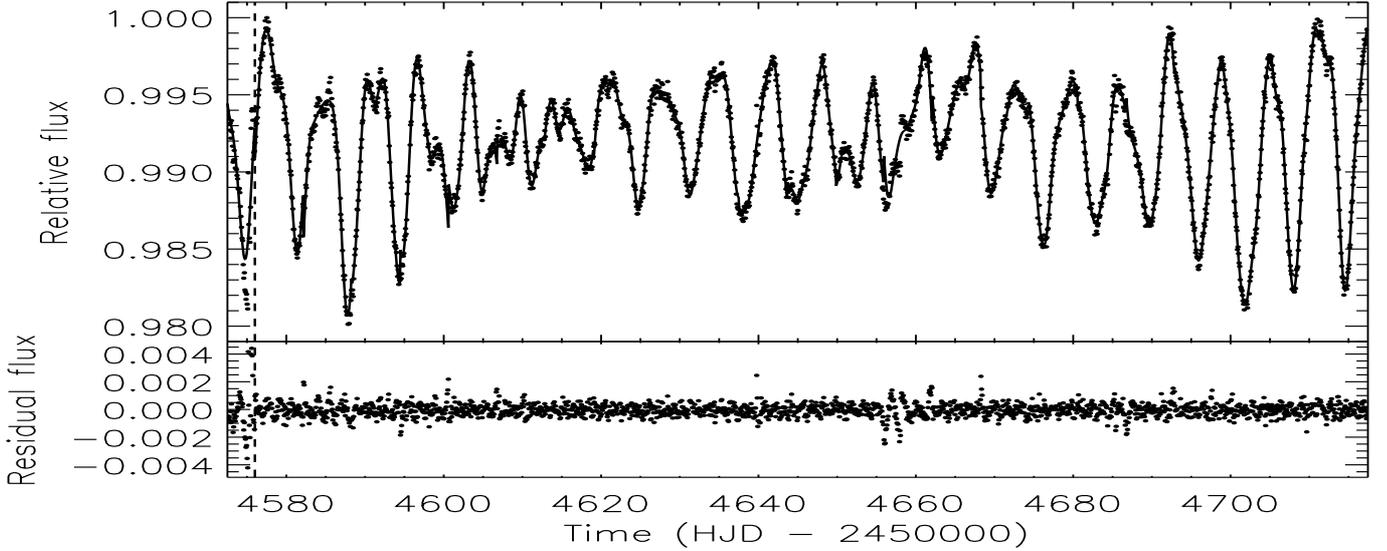}} 
\caption{{\it Upper panel:} The out-of-transit light curve of CoRoT-6 (dots) and its  ME-regularized  best fit  for a facular-to-spotted area ratio of $Q=1.5$ (solid line) during the time interval between HJD~2454572.5075 and  2454717.4519. The flux is normalized to the maximum observed flux.  {\it Lower panel:} The corresponding residuals. In both panels, the vertical dashed line marks HJD~2454576, i.e., the beginning of the interval considered for the modelling (see the text). 
}
\label{lc_bestfit}
\end{figure*}

\section{Results}
\label{results}

\subsection{Light curve model}
\label{light_curve_model}

We initially apply the model of Sect.~\ref{spotmodel} to the whole out-of-transit CoRoT-6 light curve.
Even without any regularization, the fit to the time interval between HJD~2454572.5075 and 2454576.0 is not
satisfactory owing to the remarkable flux variation on a timescale of $\sim 2$ days associated with the rapid evolution of some active regions (cf. Fig.~\ref{initial_segment}). It cannot be adequately fitted by our model which assumes that the pattern of active regions stays constant during the time interval $\Delta t_{\rm f} > 2$ days. 
Therefore, we decide to discard the first interval of the light curve, 
{ the endpoint of which is marked by the vertical dashed line in Fig.~\ref{initial_segment},} and analyse the section from 
HJD~2454576.0 to 2454717.4519 covering 141.4519 days. We consider 23 individual intervals of $\Delta t_{\rm f} = 6.15$ days which warrant a good fitting and an adequate time resolution to trace the evolution of the active regions. 
The best fit without regularization ($\lambda = 0$) has a mean  $\mu_{\rm res} = 5.739 \times 10^{-6}$ and a standard deviation of the residuals $\sigma_{0} = 4.155 \times 10^{-4}$ in relative flux units which is 1.7 times the standard error of the mean points. This is due to the flux fluctuations on a timescale of $\sim 2$ days associated with a populations of short-lived active regions that can be  only partially fitted by our approach (see  below). 
The Lagrangian multiplier $\lambda$ is iteratively adjusted until the mean of the residuals $\mu_{\rm res} = -4.962 \times 10^{-5} \simeq - \sigma_{0} / \sqrt{N}$, where $N  =  84$ is the mean number of  points in each fitted light curve interval  $\Delta t_{\rm f}$. The standard deviation of the residuals of the regularized best fit is $\sigma = 4.728 \times 10^{-4}$. 

The composite best fit to the entire light curve is shown in the upper panel of Fig.~\ref{lc_bestfit} while the residuals are plotted in the lower panel. The fit is generally good, although the residuals sometimes show significant excursions to be attributed to the rapidly evolving active regions with lifetimes of $1-3$ days that cannot be fitted by our approach. 
 In particular, during the first part of the observations up to HJD~2454576 marked by the vertical dashed line, the residuals are a factor of $\sim 4-5$ larger than during most of the subsequent interval. 

 A ME model computed from HJD~2454576.0 to 2454717.4519 with $Q=4$ shows a marginally better fit with an average $\sigma = 4.572 \times 10^{-4}$. Nevertheless, we decide to adopt $Q=1.5$ in our analysis, corresponding to the minimum of the total $\chi^{2}$ of the fit obtained with the three-spot model,  as described in Sect.~\ref{model_param}.

\subsection{Longitude distribution of active regions and stellar differential rotation}
\label{spot_model_res}

\begin{figure}[]
\centerline{
\includegraphics[height=11cm,angle=90]{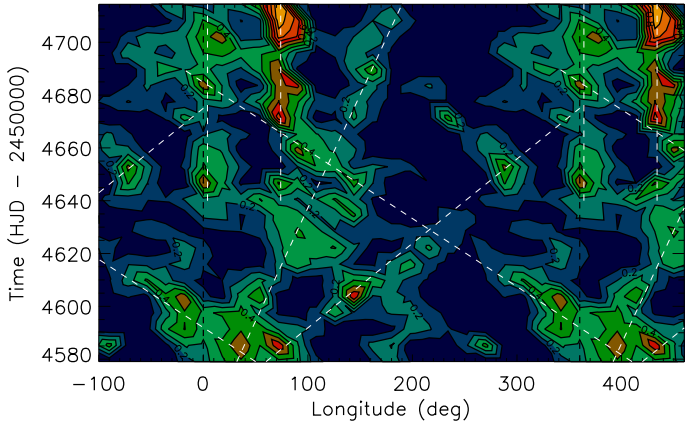}}  
\caption{Isocontours of the ratio $f/f_{\rm max}$, where $f$ is the spot covering factor and $f_{\rm max}= 0.0059$ its maximum value,  versus time and longitude for the ME models with $Q=1.5$. The two dashed black lines mark longitudes $0^{\circ}$ and $360^{\circ}$ beyond which the distributions are repeated to easily follow spot migration. The contour levels are separated by $\Delta f= 0.1 f_{\rm max}$ with yellow indicating the maximum covering factor and dark blue the minimum. The dashed white lines trace the migration of the active regions associated with  each active longitude (see the text). }
\label{synop_spot}
\end{figure}

The distribution of the spot covering factor $f$ versus longitude and time is plotted in Fig.~\ref{synop_spot}
for our ME models with $Q=1.5$. The  origin of the longitude corresponds to the sub-observer point at the epoch HJD~2454576.0 and it increases in the same direction of stellar rotation and the orbital motion of the planet. 
The lifetimes of the mapped active regions range from $\sim 6$ days, that is the time resolution of our spot modelling, to $\sim 30-40$ days,  as can be inferred from the isocontours of their filling factors in Fig.~\ref{synop_spot}. Most of the active regions grow and decay within five active longitudes that can be identified by their  uniform migration in the plot. Although there is always some degree of arbitrareness in attributing a given active region to a specific active longitude, we trace the migration of these longitudes by white dashed lines. Their migration rates range from $-3.88 \pm 0.75$ deg/day to $3.13 \pm 0.75$ deg/day. If such  different  rates correspond to the angular velocity at different latitudes on a differentially rotating star, the relative amplitude of the differential rotation is estimated to be $\Delta \Omega / \Omega = 0.123 \pm 0.020 $. This is actually a lower limit for the pole-equator angular velocity difference because we do not have information on the latitude of the active regions which are probably confined to some range as we observe in the Sun where spots appear only in the $\pm 40^{\circ}$ belt.  A consequence of this interpretation is that the intersections between the straight lines tracing the migration of different active longitudes in Fig.~\ref{synop_spot} do not necessarily imply a coincidence of the corresponding active regions because they are probably located at different latitudes.

A ME spot model computed with $Q=4$ leads to a somewhat smaller differential rotation, i.e., $\Delta \Omega / \Omega \simeq 0.09 \pm 0.02$, because the reconstructed longitudes of the active regions are more affected by the  flux modulation due to their facular component (cf. Sect.~\ref{spotmodel}). 
 Reducing the inclination $i$ of the stellar rotation axis to $70^{\circ}$ does not significantly change the amplitude of  the differential rotation. On the other hand, for $i=50^{\circ}$, several small spots  are ill-mapped making it difficult to trace the migration of  the active longitudes rotating faster than the mean period of $6.35$ days. This leads to an estimated $\Delta \Omega / \Omega$ about half the above  value.  

The amplitude of the surface differential rotation of CoRoT-6 is in agreement with the mean value found for stars of comparable effective temperature and rotation period by fitting the shape of their line profiles or by Doppler imaging techniques \citep{Barnesetal05,Reiners06}. 
For CoRoT-4, that has a slightly hotter effective temperature of 6190~K and a mean rotation period of $\sim9$ days, we estimate a lower limit for the surface differential rotation  $\Delta \Omega / \Omega = 0.057 \pm 0.015$ from the ME spot models with $Q=4.5$ \citep{Lanzaetal09b}. It may be explained by assuming that the active regions mapped on that star are closer to the equator. By comparison, the relative angular velocity difference in the $\pm 40^{\circ}$ sunspot belt amounts to $\sim 0.05$. 

\begin{figure}[!h]
\centerline{
\includegraphics[width=6cm,angle=90]{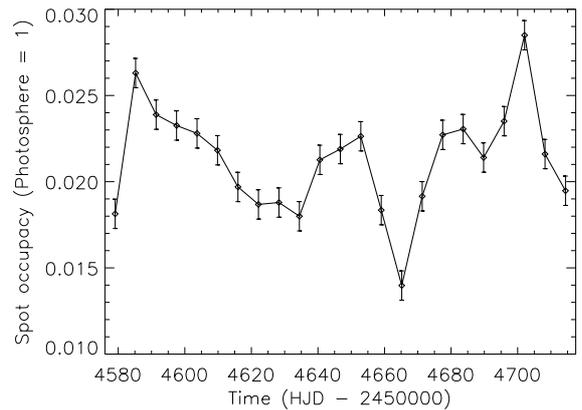}} 
\caption{{The total spotted area vs. time, as derived from our lightcurve model illustrated in Figs.~\ref{lc_bestfit} and~\ref{synop_spot}. The uncertainty of the area (3$\sigma$ errorbars) has been derived from the standard deviation of the residuals of the best fit to the light curve. }   
}
\label{total_area}
\end{figure}

The total spotted area as derived from our modelling  is plotted vs. time in Fig.~\ref{total_area}. Since we removed the long-term linear trend in the flux, the associated variation of the spotted area is not included in our model. There is no clear periodicity in the variation, although timescales ranging from $\sim 25$ to $\sim 60$ days are apparent. 

\subsection{Signatures of a possible star-planet interaction}
\label{star-planet-inter}

In the reference frame adopted for our spot modelling, which rotates with a period  $P_{\rm rot} = 6.35$ days with respect to an inertial frame, the period between subsequent crossings of a given meridian by the subplanetary point is the synodic period $P_{\rm syn}=22.25$ days, defined as $P_{\rm syn}^{-1} = P_{\rm rot}^{-1} - P_{\rm orb}^{-1}$, where the orbital period of the planet $P_{\rm orb} = 8.886$ days. 

\begin{figure}[]
\centerline{
\includegraphics[height=12cm,angle=90]{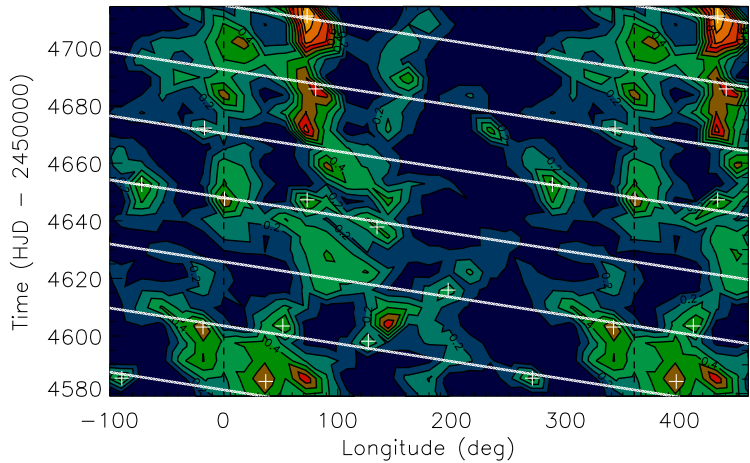}} 
\caption{Isocontours of the spot filling factor  as in Fig.~\ref{synop_spot} with white lines tracing the active  meridian lagging the subplanetary meridian by $\Delta \ell = -200^{\circ}$. White crosses indicate the relative maxima of the spot filling factor corresponding to the active regions listed in Table~\ref{spi_list}.}
\label{subplan_plot}
\end{figure}

Considering the distribution of the spot filling factor $f$ vs. time and longitude, we  find a correlation of the longitude of some active regions with the meridian located at $\Delta \ell = -200^{\circ}$ with respect to the  subplanetary longitude (cf. Fig.~\ref{subplan_plot}). The white lines overplotted on the distribution of $f$ show the motion of such a meridian vs. time with the period $P_{\rm syn}$. The active regions that reach a maximum  filling factor close to their passage across this meridian are  marked with crosses in Fig.~\ref{subplan_plot} and  listed in Table~\ref{spi_list}, where the columns from the left to the right report the time of their maximum filling factor, their  longitude $\ell_{\rm ar}$ at that time,  the filling factor per $18^{\circ}$ longitude bin $f/f_{\rm max}$, normalized to the maximum filling factor over the whole map ($f_{\rm max}=0.0059$), the distance $\delta \ell$ of the active region from the active meridian, and the probability $p$ of a chance association between them, respectively. The latter is computed according to:
\begin{equation}
p = \min \left\{ n_{\rm ar} \frac{\delta \ell}{\Delta L}, 1 \right\},
\end{equation}  
where $n_{\rm ar}$ is the number of active regions appearing during a given synodic period and $\Delta L$ the  longitude range  where stellar activity is  mainly localized. If  activity were distributed uniformly, $\Delta L = 360^{\circ}$, but in CoRoT-6 most of the active regions appear within a longitude range  $\Delta L \sim 250^{\circ}$, owing to the presence of active longitudes. Therefore, we adopt $\Delta L =250^{\circ}$ and $n_{\rm ar} = 7$ to compute the probabilities in the last column of Table~\ref{spi_list}. The adopted $n_{\rm ar}$ corresponds to the maximum number of active regions observed in one synodic period, thus yielding the maximum probability of chance association. 

The total probability $p_{\rm T}$ of a chance occurrence for the entire set of 13 active regions is:
\begin{equation}
 p_{\rm T} = \frac{\displaystyle \prod_{i} p_{i}}{\displaystyle \min \{ p_{i} \} }.  
\end{equation}
The division by $\min \{ p_{i} \}$ accounts for 
the additional degree of freedom related to the fact that the phase lag $\Delta \ell = -200^{\circ}$ between the active and the subplanetary meridians is not fixed a priori but is derived from the same set of data. In such a way, we obtain a probability of chance occurrence $p_{\rm T} = 0.0075$, indicating that the association of those active regions with the active meridian is significant. 

Spot models with a greater facular contribution, i.e., $Q=4.0$, confirm the presence of active regions associated with the orbital motion of the planet, although $\Delta \ell = -170^{\circ}$, owing to the systematic longitude shifts of the active regions associated with a greater value of $Q$. 

Since it is not possible to exclude a misalignment between the stellar spin  and the orbital angular momentum, we have  investigated the impact of a different inclination of the stellar rotation axis on the above correlation finding that the active regions listed in Table~\ref{spi_list} are well reproduced  also when $i = 70^{\circ}$. On the other hand,  only one third of them survive when $i=50^{\circ}$, which corresponds to the minimum inclination compatible with the estimated stellar parameters (cf. Sect.~\ref{model_param}). In this case, no significant evidence of  star-planet interaction is found.
\begin{table}
\noindent 
\caption{Stellar active regions associated with the active meridian.}
\begin{center}
\begin{tabular}{crcrc}
\hline
&  & & & \\
Time &  $\ell_{\rm ar}\;\;$  & $ f/f_{\rm max} $ & $\delta \ell \;\; $  & $ p $ \\
(HJD-2450000)  & (deg) & & (deg) & \\ 
 & & & & \\
\hline
& &  & & \\
    4583.68 &   36.2 & 0.563 &  75.1  & 1.000 \\
    4584.81 &  270.0 & 0.417 &  32.8  & 0.918 \\
    4597.65 &  126.2 & 0.387 &   31.2  & 0.873 \\
    4602.56 &  341.2 & 0.184 &   34.4  & 0.964 \\
    4602.93 &   51.2 & 0.452 &   41.7  & 1.000 \\
    4615.39 &  196.2 & 0.327 &   28.3 &  0.792 \\ 
    4637.29 &  133.8 & 0.328 &   39.9  & 1.000 \\
    4646.73 &   72.5 & 0.515 &   51.6  & 1.000 \\
    4647.10 &    0.0 & 0.591 &  14.8  & 0.414 \\
    4652.01 &  287.5 & 0.505 &   7.9  & 0.220 \\
    4671.64 &  342.5 & 0.335 &   4.8  & 0.135 \\
    4685.23 &   80.0 & 0.543 &  37.8  & 1.000 \\
    4710.53 &   72.5 & 0.965 &   4.0  & 0.113 \\
 & & & & \\
\hline
\label{spi_list}
\end{tabular}
\end{center}
\end{table}

\begin{figure}[!b]
\centerline{
\includegraphics[height=9cm,angle=90]{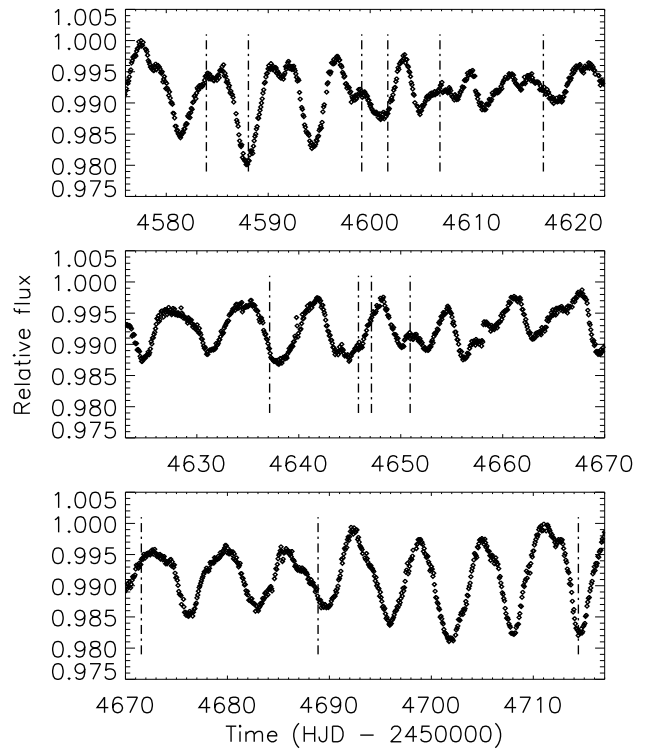}} 
\caption{The light curve of CoRoT-6 with the epochs of maximum visibility of the active regions listed in Table~\ref{spi_list} marked by vertical dash-dotted lines. }   
\label{best_visibility}
\end{figure}

In Fig.~\ref{best_visibility} we plot the light curve of CoRoT-6 and mark the epochs corresponding to the passages  of the active regions listed in Table~\ref{spi_list} across the central meridian of the stellar disc. The two remarkable active regions reaching the maximum filling factor at HJD~2454583.68 and 2454710.53 correspond to flux minima at HJD~2454588.1 and 2454714.4, respectively, while the other smaller active regions are generally associated with changes of the slope of the light curve.  Such slope changes can be attributed to the variation of their area (or filling factor) because the projection factor is constant when an active region crosses the central meridian. 

The separations between the vertical marks in Fig.~\ref{best_visibility} depend on the epochs of maximum visibility of the corresponding active regions, which coincide with the times when they cross the central meridian of the disc and are in general not simultaneous with  the maxima of their filling factors as listed in Table~\ref{spi_list}. Specifically, given that the rotation period of our reference frame is $P_{\rm rot}=6.35$ days, the mark corresponding to a given active region in Fig.~\ref{best_visibility} is shifted from the epoch of its maximum filling factor by a time interval depending on its longitude that can range up to
$\pm P_{\rm rot}$. Their different longitudes and the fact that the  active regions associated with the star-planet interaction do not appear periodically, lead to an uneven  distribution of the corresponding marks vs. time in Fig.~\ref{best_visibility}.  


\section{Discussion}

CoRoT-6 is an interesting system to study possible  interactions between the planet and its host star because its rotation  is not synchronized with the planetary orbit allowing us to separate the rotational modulation from a signal at the orbital or the synodic period. For comparison, \object{CoRoT-4} is a slightly hotter star, but it is synchronized, thus an activity signature of planetary origin is more difficult to find out. Nevertheless, a persistent active region close to the subplanetary longitude has been observed in that star during the $\sim 70$ days of CoRoT observations \citep{Lanzaetal09b}. Another interesting case is that of \object{$\tau$~Bootis}, also synchronized to the orbital motion of its hot Jupiter. In this star, \citet{Walkeretal08} found 
evidence of a persistent photospheric active region leading the subplanetary longitude by $\approx 70^{\circ}$. 

The other remarkably active star \object{CoRoT-2} shows  a short-term activity cycle with a period of $\sim 28.9$ days, as indicated by the variation of the total area of its spots, that is very close to ten synodic periods of the planet with respect to the mean stellar rotation period of 4.52 days. As a matter of fact, the synodic period of CoRoT-2 is only 2.84 days, which is shorter than the time resolution $\Delta t_{\rm f}=3.16$ days of the spot mapping performed by \citet{Lanzaetal09a} making it impossible to look for the effects found in CoRoT-6 whose synodic period is 22.25 days, i.e.,  remarkably longer than the time resolution of our mapping   ($\Delta t_{\rm f} = 6.15$ days). However, \citet{Paganoetal09} find a systematic increase of the variance of the optical flux immediately before the planetary transits in the light curve of CoRoT-2 which  suggests short-lived active regions or flaring activity close to the subplanetary longitude.

A major difficulty with the allegedly cases of magnetic star-planet interactions comes from the remarkably different behaviours observed in different stars that makes it impossible to define a common and simple phenomenology. Moreover, in the case of CoRoT-6, there is no stable periodicity in the emergence of the active regions allegedly associated with the interaction and this makes the proposed correlation  true only in a statistical sense. The lifetimes of the active regions listed in Table~\ref{spi_list} are comparable with the synodic period, thus,  sooner or later, they should cross the straight lines marking the active meridian in Fig.~\ref{subplan_plot} during their migration at the rates characteristic of the active longitudes to which they belong. Therefore, what suggests  the presence of a star-planet interaction is not just the coincidence in longitude, but the fact that the maximum filling factors of those active regions are reached very close to such a crossing. 

This can be interpreted as the result of some triggering of activity associated with the passage of the active meridian over a previously existing active region, possibly due to an enhancement of the rate of magnetic flux emergence from the subphotospheric layers where magnetic flux is  stored. In principle, two possible mechanisms can be invoked to explain this effect, respectively associated with the tidal force or with the magnetic field  perturbation in the stellar corona.

Tidal effects in CoRoT-6 are very weak because the semimajor axis of the planetary orbit is 17.9 times the radius of the star \citep{Fridlundetal10}. Specifically, the tidal synchronization time for the star is $\sim 900$ Gyr for a  tidal quality factor $Q^{\prime} = 10^{6}$, i.e., much longer than the estimated stellar age ranging between 1 and 3.3 Gyr \citep{Fridlundetal10}. Moreover, tidal effects are unlikely to be responsible for the  observed star-planet interaction because they would produce two maxima of activity on opposite hemispheres associated with the two opposite tidal bulges  instead of the suggested intermittent activity  associated with a meridian rotating with the synodic period. 

A  perturbation of the hydromagnetic dynamo  in the subsurface layers of a star induced  by a close-in planet has been considered by \citet{Lanza08}. 
The conjectured process assumes  that the reconnection  occurring in the stellar corona between the magnetic fields  of the star and the planet  generates a perturbation of the magnetic helicity that is conducted down to the stellar surface and immediately below it along magnetic field lines.  The longitude where the perturbation of the dynamo action is maximum coincides with the footpoint of the magnetic field line connecting the surface of the star to the planet. If the stellar dynamo is not axisymmetric, when an active longitude crosses that footpoint the magnetic field in the subphotospheric layers can be intensified locally, giving rise to the observed maximum of activity in the  active region. The phenomenon is not periodic because the emergence of the magnetic flux requires that the intensity of the unperturbed field be already close to the threshold for the buoyancy instability and this  occurs 
only where there is a pre-existing active region. Since active regions appear in a non-periodic fashion and are preferentially located within active longitudes  migrating at different rates, there is no well-defined  periodicity in the star-planet interaction, but only some correlation with the activity already found in the active longitudes. 

In the   model of \citet{Lanza08}, the coronal magnetic field has a non-potential configuration so its field lines are twisted and this can account for the observed lag $\Delta \ell$ between the active and the subplanetary meridians. \citet{Lanza09} proposes that the minimum energy configuration of the stellar coronal field is a linear force-free state, i.e., with the field satisfying the equation $\nabla \times {\vec B} = \alpha {\vec B}$, where $\alpha$ is uniform. In principle, a unique linear force-free configuration can be found when the magnetic field components are given at the photosphere and the angle of twist of the field lines between the surface of the star and the orbit of the planet is specified. Unfortunately, in our case we have only information on the latter quantity, so there are potentially infinite configurations leading to the observed phase lag between the active longitude and the subplanetary point. For a phase lag  $\Delta \ell = -200^{\circ}$, one of the possible configurations is obtained for $\alpha=0.085$, while the other field parameters specifying the variation of the field strength with radius have been fixed at the values $b_{0}=-1.1$ and $c_{0}=1.0$, as adopted by \citet{Lanza08}.  

The topology of the coronal magnetic field plays an important role  in the angular momentum loss of the star through  its magnetized wind. The interaction between the planet and the coronal field could modify this topology leading to a reduction of the braking rate by $\approx 30$ percent \citep[see ][ for details]{Lanza10}. Indeed, the rotation period of CoRoT-6 seems to be quite short for a star of its mass and a minimum estimated age of $\approx 1$~Gyr \citep{Fridlundetal10}. Specifically, the gyrochronology relationship of \citet{Barnes09} gives a period  of $\sim 9$ days, i.e., about 30 percent longer than observed for a star of that mass and age. The difference is significant because the age spread allowed for $P_{\rm rot} \sim 6-7$ days is about $\pm 100$ Myr for a predicted age of $\sim 500$ Myr. 
On the other hand, it is not possible to exclude that the star is remarkably younger than 1~Gyr because this lower limit was estimated from the Lithium abundance which has a weak dependence on the stellar age in that range of mass. Of course, a younger age ($\approx 400-600$ Myr) would reconcile gyrochronology with the measured rotation period. In the case of CoRoT-6, tidal effects on the evolution of the stellar spin  are unlikely, given the very weak tidal force exerted by the planet that moves on a relatively wide orbit. 

Finally, we note that possible signatures of the star-planet interaction can be searched for also in the modulation of the chromospheric line fluxes \citep[e.g.][]{Shkolniketal05,Shkolniketal08}. From the correlation among chromospheric emission,  spectral type and rotation found by \citet{Noyesetal84}, a residual  flux in the core of the Ca~II~H\&K lines $R_{\rm HK}^{\prime} \approx 2.8 $ that of the Sun can be estimated (Foing, private comm.). It could be observable if high resolution ($\ga 60000$) and sufficiently high signal-to-noise ratio ($S/N \ga 30-40$) spectra of the Ca~II H\&K region are available \citep[cf., e.g.,][]{Reboloetal89}. Since the $v \sin i$ of CoRoT-6 is $ 7.5\pm 1.0$ km~s$^{-1}$, it could be possible to derive the longitudes of the chromospheric plages by, e.g., the method proposed  by \citet{CharFoing93}, and study their possible relationship with the planetary orbital phase.

\section{Conclusions}
\label{conclusions}

We have applied a ME spot model to fit the optical light curve of the active planet-hosting star CoRoT-6 deriving maps of the distributions of its active regions vs. longitude with a time resolution of $6.15$ days during 141 days of uninterrupted CoRoT observations. Active regions are assumed to consist of cool spots and bright faculae with  solar-like contrasts. The facular-to-spotted area ratio in each active region is assumed to be constant and the best value is found to be  $Q=1.5$ with an acceptable range between 0 and $\sim 4$. This is typical of a star with an activity level higher than the Sun.

Our maps show five active longitudes where individual active regions preferentially grow and decay with typical lifetimes up to $\sim 30-40$ days. The migration of the active longitudes allows us to derive a lower limit for the surface differential rotation $\Delta \Omega / \Omega = 0.12 \pm 0.02$. 

We find  several active regions reaching a maximum of activity at a longitude lagging the subplanetary meridian by $\Delta \ell \sim  -200^{\circ}$. A spurious coincidence, owing to a chance appearance of those active regions, can be excluded with a degree of confidence greater than 99 percent. This suggests the presence of some magnetic star-planet interaction, although the correlation we found  is significant only in a statistical sense.

\begin{acknowledgements}
The authors are grateful to the Referee, Prof. Gordon Walker, for a careful reading of the manuscript and valuable comments, and to Dr. Bernard Foing for interesting discussions. 
Active star research and exoplanetary studies at INAF-Osservatorio Astrofisico di Catania and Dipartimento di Fisica e Astronomia dell'Universit\`a degli Studi di Catania 
 are funded by MIUR ({\it Ministero dell'Istruzione, dell'Universit\`a e della Ricerca}) and by {\it Regione Siciliana}, whose financial support is gratefully
acknowledged. 
This research has made use of the CoRoT Public Data Archive operated at IAS, Orsay, France, and of the ADS-CDS databases, operated at the CDS, Strasbourg, France.
\end{acknowledgements}

\end{document}